\documentclass[twocolumn]{jpsj2} %% two-column layout
\newcommand{\bk}{\boldsymbol{k}}
\newcommand{\bq}{\boldsymbol{q}}
\newcommand{\bK}{\boldsymbol{K}}
\newcommand{\id}{\mathrm{d}}
\newcommand{\qs}{q_{\mathrm{s}}}
\newcommand{\qTF}{q_{\mathrm{TF}}}
\newcommand{\Vc}{V_{\mathrm{cell}}}
\newcommand{\Tc}{T_{\mathrm{c}}}
\newcommand{\kB}{k_{\mathrm{B}}}

\newcommand{\omD}{\omega_{\mathrm{D}}}
\newcommand{\as}{a_{\mathrm{s}}}
\newcommand{\asbs}{a_{\mathrm{s}} / b_{\mathrm{s}}}

\newcommand{\tmtx}{$\mathrm{(TMTSF)}_2\mathrm{X}$}
\newcommand{\tmtp}{$\mathrm{(TMTSF)}_2\mathrm{PF}_6$}
\newcommand{\tmtc}{$\mathrm{(TMTSF)}_2\mathrm{ClO}_4$}

\newcommand{\sw}{\textit{s}-wave}
\newcommand{\pxw}{$p_x$-wave}
\newcommand{\pyw}{$p_y$-wave}
\newcommand{\dw}{\textit{d}-wave}

\def\runauthor{
Y.~{\sc Suginishi} and H.~{\sc Shimahara} 
}

\title{
Spin-Triplet Superconductivity 
Mediated by Phonons \\
in Quasi-One-Dimensional Systems 
}

\author{Yuuichi \textsc{Suginishi} and 
Hiroshi \textsc{Shimahara}
}

\inst{Department of Quantum Matter Science, 
ADSM, 
Hiroshima University, \\
Higashi-Hiroshima 739-8530, Japan
}

\abst{%Abstract goes here.  The abstract must be a single paragraph.
We investigate 
the spin-triplet superconductivity 
mediated by phonons 
in quasi-one-dimensional (Q1D) systems with open Fermi surfaces. 
We obtain the ground state phase diagrams.
It is found that spin-triplet superconductivity 
occurs for weak screening 
and strong on-site Coulomb interaction, 
even in the absence 
of any additional nonphonon pairing interactions.
We find that the nodeless spin-triplet state 
is more favorable than the spin-triplet state 
with line nodes, 
for the parameter values 
of the Q1D superconductors $\mathrm{(TMTSF)}_2\mathrm{X}$.
We also find that Q1D open Fermi surface, 
which is the specific feature of this system, 
plays an essential role in the pairing symmetry. 
We discuss the compatibility of the present results 
with the experimental results in these compounds.
}

\kword{anisotropic superconductivity, 
       spin-triplet pairing, 
       phonon-mediated pairing interactions, 
       quasi-one-dimensional systems, 
       organic superconductors, 
       $\mathsf{(TMTSF)_2X}$
       }

\begin{document}
\maketitle

\section{\label{sec:int}Introduction} %% No sections necessary for express letters, letters and short notes

The relation between the symmetry of 
the superconducting order parameter 
and the mechanism of the pairing interaction 
in exotic superconductors 
is one of the most important subjects 
in condensed matter physics. 
In this paper, 
we are interested in this subject in
quasi-one-dimensional (Q1D) systems, 
in connection with the organic superconductors 
\tmtx\ with 
$\mathrm{X} = \mathrm{PF}_6$, $\mathrm{ClO}_4$, etc.

The compound \tmtp\ exhibits superconductivity 
in the proximity of spin-density-wave (SDW) 
in the phase diagram~\cite{Jerome0}. 
Hence, at an early stage after the discovery, 
the possibility of the superconductivity 
induced by antiferromagnetic spin fluctuations was proposed~\cite{Emery1}. 
Takigawa \textit{et al}.\ 
have measured the NMR relaxation rate $T_1^{-1}$
in superconducting \tmtc~\cite{Takigawa1}. 
Their results suggest 
the anisotropic superconducting order parameter 
with line nodes on the Fermi surface 
as Hasegawa and Fukuyama argued~\cite{Hasegawa1}. 
These experimental and theoretical results on $T_1^{-1}$ 
seem to be consistent with the mechanism of 
the superconductivity induced by the antiferromagnetic spin fluctuations. 
In this mechanism, 
the phase diagram of SDW and \dw\ superconductivity, 
whose order parameter has line nodes, 
has been reproduced semiquantitatively~\cite{Shimahara1,footnote2}.

However, recent experimental 
results seem contradictory to one another with respect to 
the symmetry of the order parameter. 
Belin and Behnia obtained 
a temperature dependence of the thermal conductivity 
in \tmtc\ 
that indicated a nodeless superconducting order parameter~\cite{Belin1}.
Their result seems inconsistent with the NMR result 
mentioned above~\cite{Takigawa1}, 
which indicates the line nodes.
As a possible consistent explanation of these results of the NMR and 
the thermal conductivity, 
a nodeless \dw\ superconductivity mediated by antiferromagnetic 
spin fluctuations has been proposed~\cite{Shimahara2}. 
Furthermore, 
in \tmtp\ at $P \simeq 6$~kbar 
and \tmtc, 
the upper critical field $H_{\mathrm{c2}}(T)$ 
seems to exceed 
the Pauli paramagnetic limit $H_{\mathrm{P}}$ 
at low temperature~\cite{Lee2,Oh1}.
From this result of $H_{\mathrm{c2}}$, 
several authors have discussed 
possibilities of spin-triplet superconductivity~\cite{Lebed1,Vaccarella1}.
However, 
there has not been any observation of a reentrant superconducting phase, 
which should occur in spin-triplet superconductors 
at higher field~\cite{Lebed2,Dupuis1}. 
The experimental result $H_{\mathrm{c2}} > H_{\mathrm{P}}$ 
can be explained by 
the Fulde-Ferrell-Larkin-Ovchinnikov 
(FFLO or LOFF) state~\cite{Lebed1,Miyazaki1,Shimahara7} and 
strong coupling effects.
Recently, Lee \textit{et al}.\ have measured a Knight shift 
in \tmtp\ at $P \simeq 0.65$~GPa, 
the absence of which strongly suggests spin-triplet pairing~\cite{Lee1}.

Here, we shall briefly summarize the experimental results 
and pairing symmetries. 
The full-gap state without any alternation of the sign, 
which we call \sw\ state, 
is consistent with the thermal conductivity, 
but could not explain the absence of the Hebel-Slichter peak of 
$T_1^{-1}$ at all~\cite{Takigawa1}, 
as well as the power low behavior and 
the Knight shift~\cite{Lee1}. 
It is also difficult to explain the phase diagram. 
Therefore, the \sw\ state is most unlikely, 
at least at the present. 
The singlet state with four line nodes, which we call the \dw\ state, 
appeared the most likely at an early stage from the phase diagram 
and the NMR results, 
but it could not explain the recent Knight shift results, 
unless any high-field triplet state is assumed. 
The thermal conductivity may be explained even in the \dw\ state 
with anion order~\cite{Shimahara2}. 
The spin-triplet states are compatible 
with the absence of the Knight shift.
For the open Fermi surface, 
both the full gap spin-triplet state without any line node 
and that with two line nodes 
at $k_y = 0$ are possible. 
We call the former and latter states the \pxw\ 
and \pyw\ states, respectively. 
However, the data of the NMR~$T_1^{-1}$ and the thermal conductivity 
are difficult to be explained consistently. 
The \pxw\ state could explain the thermal conductivity, 
but could not explain the temperature dependence of 
$T_1^{-1} \propto T^3$ for $\Tc / 2 \lesssim T \lesssim \Tc$. 
The coherence peak becomes smaller than 
that of the \sw\ state~\cite{Hasegawa1}, 
but there has not been any satisfactory theoretical fitting at the present. 
The \pyw\ state could explain the line node, 
but it is difficult to reproduce the thermal conductivity. 
In the $p_x$ and \pyw\ states, the pressure dependence of $\Tc$ 
does not appear to be easily explained as in the \dw\ state. 
Therefore, there is not any perfect explanation consistent 
for all experimental results at the present.

Many theoretical studies have also been presented 
in order to clarify 
the mechanism of superconductivity 
in these compounds.
In Q1D Hubbard models, 
consistently with the early result in the random phase approximation 
(RPA)~\cite{Shimahara1}, 
the fluctuation exchange (FLEX) approximation
by Kino and Kontani~\cite{Kino1}
and the third order perturbation theory 
by Nomura and Yamada~\cite{Nomura1}
have showed that \dw\ superconductivity is favorable. 
Spin-triplet \textit{f}-wave-like pairing 
was also examined by Kuroki \textit{et al}.~\cite{Kuroki1} 
and by Tanaka and Kuroki~\cite{Tanaka1}. 
There might be a consistent explanation 
on the basis of \textit{f}-wave pairing, 
although \textit{f}-wave pairing interactions are usually weak, 
because the \textit{f}-wave states oscillate many times 
in the momentum space.
Recently, Kohmoto and Sato have discussed 
the possibility of nodeless spin-triplet superconductivity 
in a Q1D system with electron-phonon interactions 
and antiferromagnetic spin fluctuations~\cite{Kohmoto1}.

The anisotropic pairing does not necessarily mean the nonphonon 
mechanism of the pairing interactions. 
Anisotropic pairing interactions mediated by phonons 
have been proposed by several authors. 
Foulkes and Gyorffy argued that \textit{p}-wave pairing 
is favorable in metals such as Rh, W and Pd, 
if the short range Coulomb interaction suppresses 
the \sw\ pairing~\cite{Foulkes1}. 
In connection with high-$\Tc$ cuprates, 
Abrikosov obtained an anisotropic \sw\ order parameter
in a model of the phonon-mediated interaction 
with weak Coulomb screening ~\cite{Abrikosov2,Abrikosov4}.
Moreover, Friedel and Kohmoto, 
and Chang \textit{et al}.\ have shown that \dw\ superconductivity 
is induced in a similar model with additional interactions 
mediated by antiferromagnetic spin fluctuations~\cite{Friedel1,Chang1}. 
Varelogiannis has also studied phonon-mediated \dw\ superconductivity 
in $\kappa$-BEDT-TTF compounds~\cite{Varelo1}. 
More recently, 
one of the authors and Kohmoto 
proposed that anisotropic superconductivity 
can be induced by the phonon-mediated interaction between electrons 
in ferromagnetic compounds, 
in connection with $\mathrm{UGe}_2$~\cite{Shimahara4}. 
They also found that 
the anisotropic pairing interactions are enhanced 
in  layered compounds with a large layer spacing~\cite{Shimahara3}. 
In their paper, 
they discussed that spin-triplet superconductivity 
in the compound $\mathrm{Sr}_2 \mathrm{RuO}_4$ 
and the organic superconductors may be induced 
by the phonon-mediated interaction.

In the compound \tmtc, very large isotope shift of $\Tc$ 
was observed~\cite{Schwenk1}, 
which suggests the large contribution 
of the phonon-mediated interaction 
to the superconductivity.
The large isotope shift may be explained by 
considering the spin fluctuations~\cite{Shimahara6}.

In this paper, 
we investigate the pairing symmetry 
of Q1D systems within the phonon-mediated pairing interactions. 
Recently, 
one of the authors have derived effective Hamiltonians
of anisotropic superconductors 
from a model of electron-phonon systems 
with coexisting short-range 
and long-range Coulomb interactions~\cite{Shimahara5}.
We apply these models to the Q1D system.
In particular, 
we examine the phonon-mediated pairing interactions 
with screened electron-phonon interactions.
Since we are motivated by the above controversy 
in the Q1D organic compounds \tmtx, 
we are particularly interested in which pairing symmetry is favored 
in the phonon mechanism for the parameter values of these compounds. 
However, the range of our study is not limited to these compounds. 
We are mainly interested in the effects of the open Fermi surface, 
which is one of the specific features of the Q1D systems. 
We show that nodeless spin-triplet state (\pxw\ state) 
is more favorable than the \pyw\ state and the \dw\ state 
in the phonon mechanism.

This paper is constructed as follows: 
In the next section 
we introduce the formulation 
and examine the anisotropic superconductivity mediated 
by phonons in Q1D systems.
In \S~\ref{sec:res}, 
we show the results of the numerical calculation 
and ground state phase diagrams.
The last section devoted to the summary and brief discussion.

\section{\label{sec:for}Formalism}

\subsection{\label{sec:fs1}
            Effective pairing interactions}

First, we consider a conventional form of pairing interactions 
mediated by phonons 
\begin{equation} \label{eq:2101}
    V_{\mathrm{eff}} (\bq,\omega + \mathrm{i}\delta) 
         = - g_0 
         \frac{\qs^2}{|\bq|^2 + \qs^2} 
         \frac{[\omega(\bq)]^2}
              {\omega^2 - [\omega(\bq)]^2} ,
\end{equation}
where $\qs^{-1}$ and $\omega(\bq)$ 
denote the range of the pairing interaction 
and the renormalized phonon dispersion energy, 
respectively. 
We do not consider the detail of the mechanism 
of the electron-ion interaction, 
but only assume the range of the pairing interaction $\qs^{-1}$.
In this sense, 
our model is phenomenological.
This simplification is justified 
when we are interested in a behavior of the pairing interaction 
in a length scale larger than the lattice constant.
For shorter length scale comparable to the lattice constant, 
the lattice anisotropy also contributes to the anisotropy of 
the electron-phonon interactions. 
However, we neglect it for simplicity, 
since we are mainly interested in the specific feature of 
the Q1D open Fermi surface, 
which is taken into account 
through the momentum dependence in $V_{\mathrm{eff}}$ 
by setting $\bq = \bk - \bk'$, 
where $\bk$ and $\bk'$ denote momenta on the Q1D Fermi surface. 
As is referred in ref.~\citen{Schrieffer1}, 
near the Fermi surface 
where $| \omega | < \omega(\bq ) $, 
the interaction becomes attractive 
due to the overscreening effect.
Except this, 
we ignore the dynamical effects in this paper.
Within the weak coupling theory, 
the frequency dependence of the pairing interaction 
$V_{\mathrm{eff}}(\bq,\omega + \mathrm{i}\delta)$ is simplified as
\begin{equation} \label{eq:2102}
  V_{\mathrm{eff}}(\bq) 
        = -g \frac{\qs^2}{|\bq|^2 + \qs^2} ,
\end{equation}
for $|\xi_{\bk}|, |\xi_{\bk'}| < \omD$, 
and $V_{\mathrm{eff}}(\bq) = 0$ otherwise, 
where $\omD$ and $\xi_{\bk}$ denote 
the Debye frequency and the electron energy 
measured from the Fermi energy, respectively. 
We assume Q1D band structure with the dispersion relation 
\begin{equation} \label{eq:dis}
\xi_{\bk} = -2 t_a \cos (k_x \as) 
          -2 t_b \cos(k_y b_\mathrm{s}) 
          - \mu ,
\end{equation}
where $\mu$, $\as$ and $b_\mathrm{s}$ 
denote the chemical potential, 
the intermolecular distances 
along the \textit{a} and \textit{b} directions, 
respectively.
In eq.~(\ref{eq:2102}), we have introduced the effective coupling constant $g$, 
which depends on the phonon dispersion. 
Since it is difficult to estimate $g_0$ and $g$ from the first principle, 
we regard $g$ as a parameter in this paper. 
Equation~(\ref{eq:2102}) has been studied by many authors 
\cite{Abrikosov2,Abrikosov4,Friedel1,Chang1,Shimahara4,Shimahara3}.

In the application of eq.~(\ref{eq:2102}) to the lattice system, 
it is more precise to replace $|\bq|$ 
with $\min_{\bK}|\bq - \bK|$~\cite{Shimahara5}. 
The effective interaction of eq.~(\ref{eq:2102}) has 
the peak around $\bq = 0$. 
Actually, the same structures should exist near $\bq = \bK$, 
where $\bK$ denotes the reciprocal lattice vector. 
Hence, we obtain the effective Hamiltonian 
\begin{equation} \label{eq:Veff_model_a}
     V_{\mathrm{eff}}(\bq) 
          = - g \, \max_{\bK} 
              \frac{\qs^2}{|\bq - \bK|^2 + \qs^2} . 
\end{equation}
We refer to this Hamiltonian 
as model~(a) in this paper.

Secondly, we introduce a model to examine the effect of 
the corrections due to the charge fluctuations. 
Within the RPA~\cite{Shimahara4,Shimahara5}, 
the effective interaction is written as 
\begin{equation} \label{eq:Veff_model_b}
      V_{\mathrm{eff}}(\bq) 
           = -g \max_{\bK} 
                \frac{\qs^2}
                     {|\bq - \bK |^2 + \qs^2 \chi_0(\bq)/N(0)} . 
\end{equation}
We refer to the Hamiltonian of 
eq.~(\ref{eq:Veff_model_b}) as model~(b). 
Here, $\chi_0(\bq)$ denotes the static free susceptibility 
\begin{equation} \label{eq:susc}
\chi_0(\bq) = \frac{1}{N}\sum_{\bk}
   \frac{f(\xi_{\bk}) - f(\xi_{\bk+\bq})}{\xi_{\bk + \bq} - \xi_{\bk}}, 
\end{equation}
where $f(x) = 1/ (\mathrm{e}^{x /\kB T} + 1)$.
In the original form of this model, 
we must set $\qs = \qTF$.
However, we extend the model by replacing 
$\qTF^{-1}$ with a correct screening length $\qs^{-1}$ 
so that it includes model~(a) with $\qs \ne \qTF$ 
as a limiting case 
$\chi_0(\bq) / N(0) \approx 1$.
This replacement of $\qTF$ by $\qs$ 
corresponds to taking into account the vertex correction 
$\Gamma(\bk, \bk', \bq)$ of $v_{\bq}$ 
by an average quantity $\bar\Gamma$ with 
$\qs^2 = \bar\Gamma \qTF^2$.

Thirdly, we introduce a model which includes the corrections 
due to the short-range Coulomb interaction $U$ 
other than the direct repulsion. 
Since the details of the derivation are presented 
in the previous paper~\cite{Shimahara5}, 
we shall describe only the outline here. 
We recall that the pairing interaction mediated by phonons 
should be proportional to ${M_{\bq}^2}/{\kappa_v(\bq)}$ 
in the absence of $U$, 
where $M_{\bq}$ and $\kappa_v(\bq)$ denote 
the electron-ion coupling constant 
and the dielectric function 
due to the long-range Coulomb interaction $v_{\bq}$, respectively. 
The electron-ion interactions are, more or less, 
the interactions between charges, 
irrespectively of their details. 
Therefore, in the presence of $U$, 
it is plausible to replace $M_{\bq}$ with $M_{\bq}/\kappa_U(\bq)$, 
where $\kappa_U(\bq)$ denotes the dielectric function due to $U$. 
Furthermore, the polarization function $\chi_0(\bq)$ 
in the dielectric function $\kappa_v(\bq)$ should be 
replaced by that with a correction due to $U$, 
which we write $\chi_U(\bq)$. 
In the RPA with respect to $U$, we have 
$\kappa_U(\bq) = 1 + U \chi_0(\bq)$ and 
\begin{equation} \label{eq:chiUdef}
     \chi_U(\bq) 
     \equiv \frac{\chi_0(\bq)}{1 + U \chi_0(\bq)} 
     =      \frac{\chi_0(\bq)}{\kappa_U(\bq)} .
\end{equation}
Therefore, eq.~(\ref{eq:Veff_model_b}) is modified as 
\begin{equation} \label{eq:Veff_model_c}
      V_{\mathrm{eff}}(\bq) 
           = -  \frac{g}{[\kappa_U(\bq)]^2}
                \max_{\bK} 
                \frac{\qs^2}
                     {|\bq - \bK |^2 + \qs^2 \chi_U(\bq)/N(0)} , 
\end{equation}
which we call model~(c) in this paper.

The range of the effective interaction $\qs^{-1}$ is 
on the order of the screening length 
in eq.~(\ref{eq:Veff_model_a}), 
because it should reflect the range of the Coulomb interaction 
between electrons and ions. 
In fact, if we adopt the Thomas-Fermi approximation, 
we obtain the effective Hamiltonian of the same form as 
eq.~(\ref{eq:Veff_model_a}) with $\qs = \qTF$. 
However, 
we should be careful when we use the Thomas-Fermi approximation 
for quantitative purpose on the screening length. 
For example, in the \tmtx\ systems, 
the Thomas-Fermi screening length is estimated as 
$\qTF^{-1} \simeq 1.05 $~\AA, 
since $t_a \approx 0.25$~eV and 
$\Vc \approx 357.2$~\AA$^3$, 
where $\Vc$ denotes the unit cell volume~\cite{Ishiguro1,footnote1}. 
Hence, $\qTF^{-1}$ becomes much smaller than 
the intermolecular distance $a_{\mathrm{s}} = 3.649$~\AA\ 
in the $a$ direction. 
However, it is obviously underestimation 
due to the Thomas-Fermi approximation. 
In practice, 
we could not apply the Thomas-Fermi approximation to 
phenomena of such a short length scale. 
In the real materials, 
the screening charges could not exist so densely 
within the distance shorter than 
the intermolecular distance $a_{\mathrm{s}}$. 
Therefore, 
for the physical reason, 
it is more appropriate to assume $\qs^{-1} \gtrsim a_{\mathrm{s}}$ 
rather than to put $q_{\mathrm{s}}^{-1} = \qTF^{-1}$ 
in the application to the \tmtx\ systems.

\subsection{\label{sec:fs2}Gap equation}

The gap equation of superconductivity is expressed as 
\begin{equation} \label{eq:2201}
\Delta(\bk) = -\frac{1}{N}
                             \sum_{\bk'} V(\bk-\bk') 
                             \frac{\tanh [E_{\bk'} / 2 \kB T]}{2E_{\bk'}}
                             \Delta(\bk') ,
\end{equation}
where we have defined
$E_{\bk'} \equiv \sqrt{\xi_{\bk'}^2 + | \Delta(\bk') |^2}$, 
and $N$ denotes the number of the lattice sites.
In this gap equation (\ref{eq:2201}), 
the summation over $\bk'$ in the right-hand side 
is restricted to $| \xi_{\bk'} | < \omD$ 
near the Fermi surface.
In the numerical calculations, 
we neglect the temperature dependence on $\chi_0(\bq)$ 
since $\kB T \ll t_a$. 
Further, we consider only intra-layer pairing and 
omit the $k_z$-dependence in $\Delta(\bk)$ in this paper. 
Therefore, 
$k_z$ integral of $V_{\mathrm{eff}}(\bq)$ appears in the gap equation. 
However, we have confirmed that 
the $k_z$ integral of $V_{\mathrm{eff}}(\bq)$ has a peak around $q_x = q_y = 0$ 
which can be fit in practice by a Lorenzian function with $\qs$ adjusted. 
Hence, we neglect $q_z$-dependence in $V_{\mathrm{eff}}(\bq)$ for simplicity.

Now, 
we consider the quasi-one-dimensional (Q1D) systems.
Once we have obtained the gap equation, 
the inter-plane electron hopping energy $t_c$ is negligible, 
while it justifies the BCS mean field approximation 
for sufficiently low temperatures.
Near the Fermi surface, we put $\bk \simeq (\pm k_{\mathrm{F}x}(k_y), k_y)$, 
where the signs $+$ and $ - $ correspond to the areas of 
the Fermi surface of $k_x > 0$ and $k_x < 0$, 
respectively.
Here, $k_{\mathrm{F}x}(k_y)$ 
is the Fermi wave vector of the $k_x$ direction at $k_y$.
Then, we can write 
\begin{align} \label{eq:2202}
\begin{split}
V(\bk - \bk') &\simeq V(k_{\mathrm{F}x}(k_y) 
     \mp k_{\mathrm{F}x}(k_y'),k_y - k_y') \\
&\equiv \begin{cases}
       V^{(++)}(k_y,k_y') = V^{(--)}(k_y,k_y') \\
       V^{(+-)}(k_y,k_y') = V^{(-+)}(k_y,k_y')
       \end{cases} 
\end{split}
\end{align}
and $\Delta(\bk) \simeq \Delta^{({\mathrm{sign}}(k_x))}(k_y)$.
In the weak-coupling limit, 
one readily finds that
the gap equation (\ref{eq:2201}) 
near the superconducting transition temperature 
$\Tc$ takes the form 
\begin{align} \label{eq:2204}
\begin{split}
&\begin{pmatrix}
\Delta^{(+)}(k_y) \\
\Delta^{(-)}(k_y)
\end{pmatrix}
   = -\frac{1}{\lambda} 
   \int_{-\pi/b}^{\pi/b} \frac{b \, \id k_y'}{2 \pi} \rho(k_y') \\ 
   & \quad \times
   \begin{pmatrix}
   V^{(++)}(k_y,k_y') & V^{(+-)}(k_y,k_y') \\
   V^{(-+)}(k_y,k_y') & V^{(--)}(k_y,k_y')
   \end{pmatrix} 
      \begin{pmatrix}
      \Delta^{(+)}(k_y') \\
      \Delta^{(-)}(k_y')
      \end{pmatrix} ,
\end{split}
\end{align}
where we have defined the density of states 
$\rho(k_y') = 1 / 4\pi t_a\sin[k_{\mathrm{F}x}(k_y')a]$ 
and 
\begin{equation} \label{eq:2206}
\frac{1}{\lambda} =\ln 
   \left[ \frac{2 \mathrm{e}^{\gamma}\omD}{\pi \kB \Tc} \right] 
\end{equation}
with the E\"{u}ler constant
$\gamma = 0.57721 \cdots$.

This matrix equation (\ref{eq:2204}) 
is easily diagonalized as
\begin{equation} \label{eq:2207}
\hat{\Delta}^D(k_y) = -\frac{1}{\lambda}
   \int_{-\pi/b}^{\pi/b} \frac{b \, \id k_y'}{2 \pi} 
   \: \hat{V}^D(k_y,k_y') \: \hat{\Delta}^D(k_y') 
\end{equation}
with
\begin{align} \label{eq:2208}
\begin{split}
\hat{\Delta}^D(k_y) &=
   \sqrt{\rho(k_y)}
   \begin{pmatrix}
   \Delta^{(+)}(k_y) + \Delta^{(-)}(k_y) \\
   \Delta^{(+)}(k_y) - \Delta^{(-)}(k_y)
   \end{pmatrix} \\
   &\equiv
   \begin{pmatrix}
   \tilde{\Delta}^{\mathrm{s}}(k_y) \\
   \tilde{\Delta}^{\mathrm{a}}(k_y)
   \end{pmatrix} 
\end{split}
\end{align}
and 
\begin{equation} \label{eq:2209}
\hat{V}^D(k_y,k_y') 
   \equiv
   \begin{pmatrix}
   \tilde{V}^{(++)} + \tilde{V}^{(+-)} & 0 \\
   0 & \tilde{V}^{(++)} - \tilde{V}^{(+-)}
   \end{pmatrix} ,
\end{equation}
where we have defined 
\begin{equation}
\tilde{V}^{(ss')}(k_y,k_y') 
\equiv \sqrt{\rho(k_y^{})}
V^{(ss')}(k_y,k_y') \sqrt{\rho(k_y')} .
\end{equation}
We expand the effective interaction 
and the order parameter on the Fermi surface as 
\begin{align} \label{eq:2210}
\begin{split}
\tilde{V}^{(ss')}(k_y,k_y') 
   = \sum_{m = 0}^\infty \sum_{n = 0}^\infty 
   &\left[
   V_{mn}^{(ss')} \: \gamma_m(k_y) \: \gamma_n(k_y') \right. \\
  &\left. + \bar{V}_{mn}^{(ss')} \: \bar{\gamma}_m(k_y) \: \bar{\gamma}_n(k_y')
   \right] 
\end{split}
\end{align}
and 
\begin{equation} \label{eq:2211}
\tilde{\Delta}^\alpha(k_y) 
   =  \sum_{m=0}^\infty \left[
   \Delta_m^\alpha \: \gamma_m(k_y) 
   + \bar{\Delta}_m^\alpha \: \bar{\gamma}_m(k_y)
   \right] 
\end{equation}
with $\alpha$ = s or a, 
which correspond to symmetric and antisymmetric states 
with respect to $k_x$, respectively.
Here, we have defined
$\gamma_m(k_y) = n_m\cos(mk_yb) $ and 
$\bar{\gamma}_m(k_y) = n_m\sin(mk_yb) $, 
and the normalization factors 
$n_m = \sqrt{2}$ for $m \ne 0$ 
and $n_m = 1$ for $m = 0$.
The expansion factors are given by 
\begin{align} \label{eq:2213}
\begin{split}
V_{mn}^{(ss')} &
   = \int_{-\pi/b}^{\pi/b} \frac{b \, \id k_y}{2 \pi} 
     \int_{-\pi/b}^{\pi/b} \frac{b \, \id k_y'}{2 \pi} \\
   & \hspace{15mm} \times \gamma_m(k_y) 
   \: \tilde{V}^{\mathrm{(ss')}}(k_y,k_y') 
   \: \gamma_n(k_y') \:, \\
\bar{V}_{mn}^{(ss')} &
   = \int_{-\pi/b}^{\pi/b} \frac{b \, \id k_y}{2 \pi} 
     \int_{-\pi/b}^{\pi/b} \frac{b \, \id k_y'}{2 \pi} \\
   & \hspace{15mm} \times \bar{\gamma}_m(k_y)
   \: \tilde{V}^{\mathrm{(ss')}}(k_y,k_y')
   \: \bar{\gamma}_n(k_y') .
\end{split}
\end{align}
By inserting eqs.~(\ref{eq:2210}) and (\ref{eq:2211}) 
into the gap equation (\ref{eq:2207}),  
we obtain matrix equations
\begin{align} \label{eq:2214}
\begin{split}
\sum_{n=0}^{\infty} \lambda_{mn}^\alpha \Delta_n^\alpha 
   &=\lambda \Delta_m^\alpha , \\
\sum_{n=0}^{\infty} \bar{\lambda}_{mn}^\alpha \bar{\Delta}_n^\alpha 
   &=\lambda \bar{\Delta}_m^\alpha ,
\end{split}
\end{align}
where 
\begin{align} \label{eq:2215}
\begin{split}
 \lambda_{mn}^\mathrm{s}
&= -\bigl(V_{mn}^{(++)} 
+ V_{mn}^{(+-)}\bigr) , \\ 
 \lambda_{mn}^\mathrm{a}
&= -\bigl(V_{mn}^{(++)} 
- V_{mn}^{(+-)}\bigr) , \\ 
 \bar{\lambda}_{mn}^\mathrm{s}
&= -\bigl(\bar{V}_{mn}^{(++)} 
+ \bar{V}_{mn}^{(+-)}\bigr) , \\ 
 \bar{\lambda}_{mn}^\mathrm{a}
&= -\bigl(\bar{V}_{mn}^{(++)} 
- \bar{V}_{mn}^{(+-)}\bigr) .
\end{split}
\end{align}
We summarize our notation 
in Table \ref{tab:01}.
\begin{table}
\caption{\label{tab:01} The classification 
of the possible pairing states 
and relevant components of the gap functions.}
\begin{center}
\begin{tabular}{rr ll} \hline \hline
&  & $\gamma_m (k_y)$ & $\bar{\gamma}_m(k_y)$ \\
[2pt]
\hline
\\ [-5pt]
$\alpha = \mathrm{s} $&  & $\Delta_0^\mathrm{s}$ : \sw\ state  
& $\bar{\Delta}_1^\mathrm{s}$ : \pyw\ state \\
&  & $\Delta_1^\mathrm{s}$ : \dw\ state & \\
\\ [-5pt]
$\alpha = \mathrm{a} $ 
&  & $\Delta_0^\mathrm{a}$ : \pxw\ state 
& $\bar{\Delta}_1^\mathrm{a}$ : $d_{xy}$-wave state \\
[2pt]
\hline
\hline
\end{tabular}
\end{center}
\end{table}

With the maximum eigenvalue $\lambda$, 
the superconducting transition temperature $\Tc$ is given by 
\begin{equation} \label{eq:2217}
\kB \Tc = \frac{2 \mathrm{e}^\gamma}{\pi}\omD \exp\left[-1/\lambda\right] ,
\end{equation}
for $\lambda > 0$.

Here, 
we note that 
the Coulomb pseudo-potential $\mu_{\mathrm{C}}^\ast$ defined by
\begin{equation} \label{eq:2218}
\mu_{\mathrm{C}}^\ast 
   = \cfrac{UN(0)}{1 + UN(0) 
      \ln \left( W_{\mathrm{C}} / \omD \right)} 
\end{equation}
must be subtracted from the eigenvalue $\lambda$ for 
the \sw\ state obtained above\cite{Shimahara3}.
Here, 
$W_{\mathrm{C}}$ is the cutoff energy on the order of the bandwidth. 
It is obvious that 
the on-site Coulomb repulsive interaction $U$
is not included in models~(a)-(c), 
except the corrections due to $U$ 
in the electron-phonon interaction 
and the screening function in model~(c).
However, 
it reduces only the \sw\ pairing interaction, 
because of the symmetry.
For example, 
in \tmtx, 
it is estimated as $U \simeq 1.5 t_a$ 
from the SDW transition temperature 
at $t_b = 0.1 t_a$~\cite{Shimahara1}.

At this point, 
we have three parameters, 
the effective electron-phonon coupling constant $g$, 
the screening length $\qs^{-1}$ 
and the on-site Coulomb interaction $U$.
Furthermore, 
we shall consider 
the electron number $n$ per a site and a spin, 
the transfer integral $t_b$ 
and the ratio $\asbs$ 
as given parameters within the Q1D systems 
with open Fermi surfaces. 
Here, 
it is easily verified 
by exchanging particle and hole 
that the results for $n$ are 
equivalent to those for $n -1$.
For example, $n = 1/4$ and $n = 3/4$ 
give the same results below.

\section{\label{sec:res}Results}

\begin{fullfigure}[t]
\begin{center}
\begin{tabular}{ccc}
\includegraphics[width=54mm,keepaspectratio]{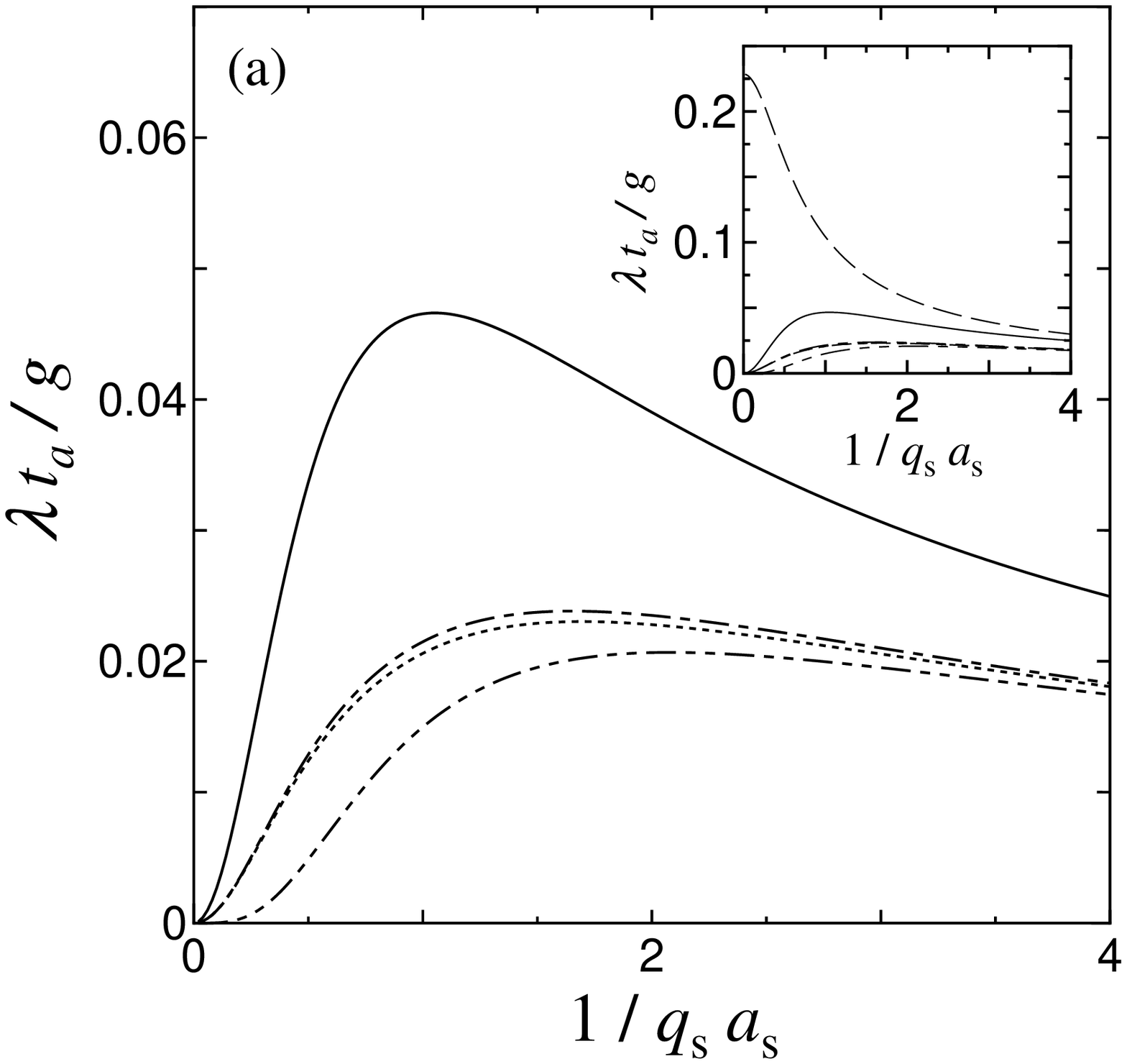} &
\includegraphics[width=54mm,keepaspectratio]{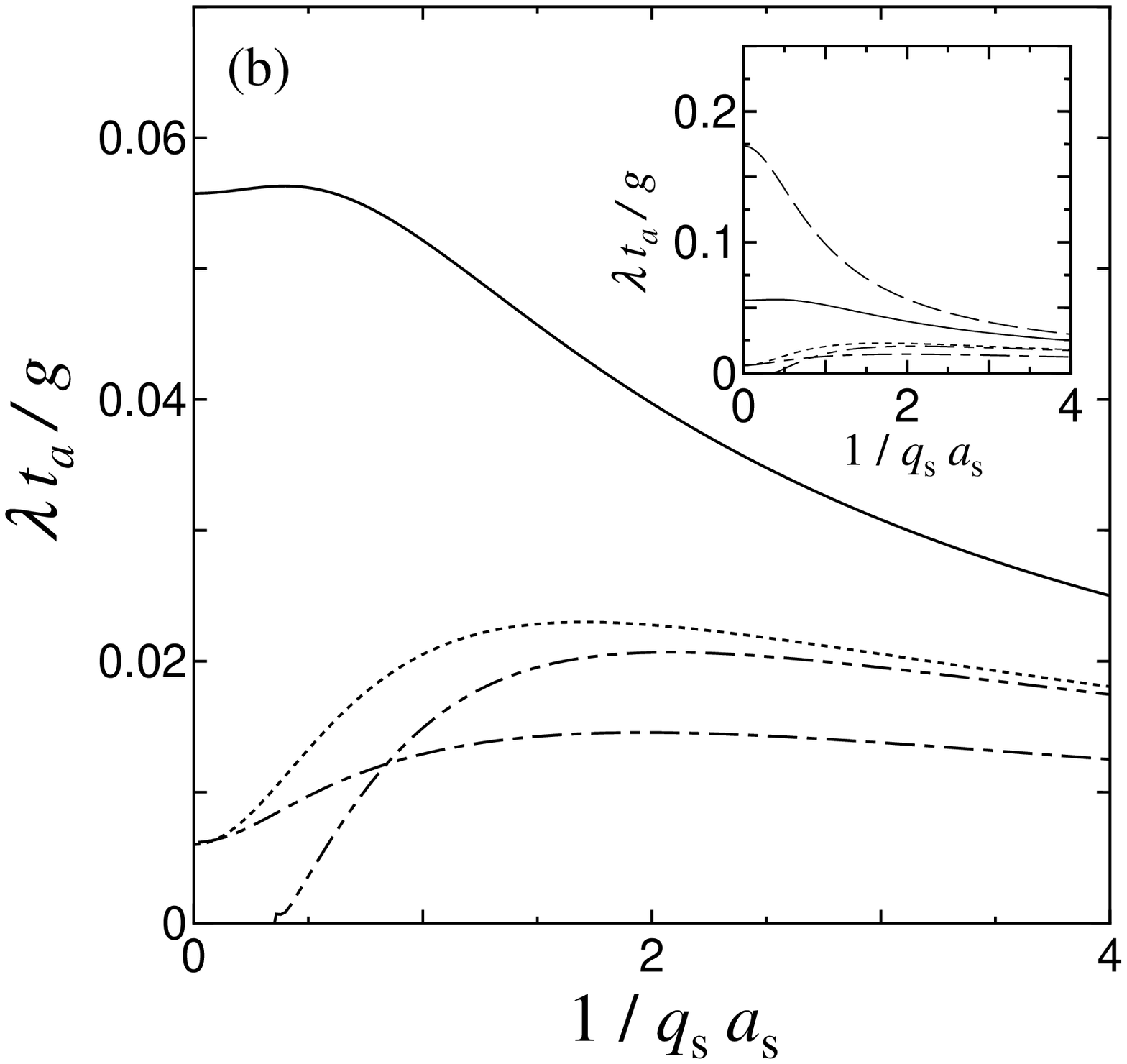} &
\includegraphics[width=54mm,keepaspectratio]{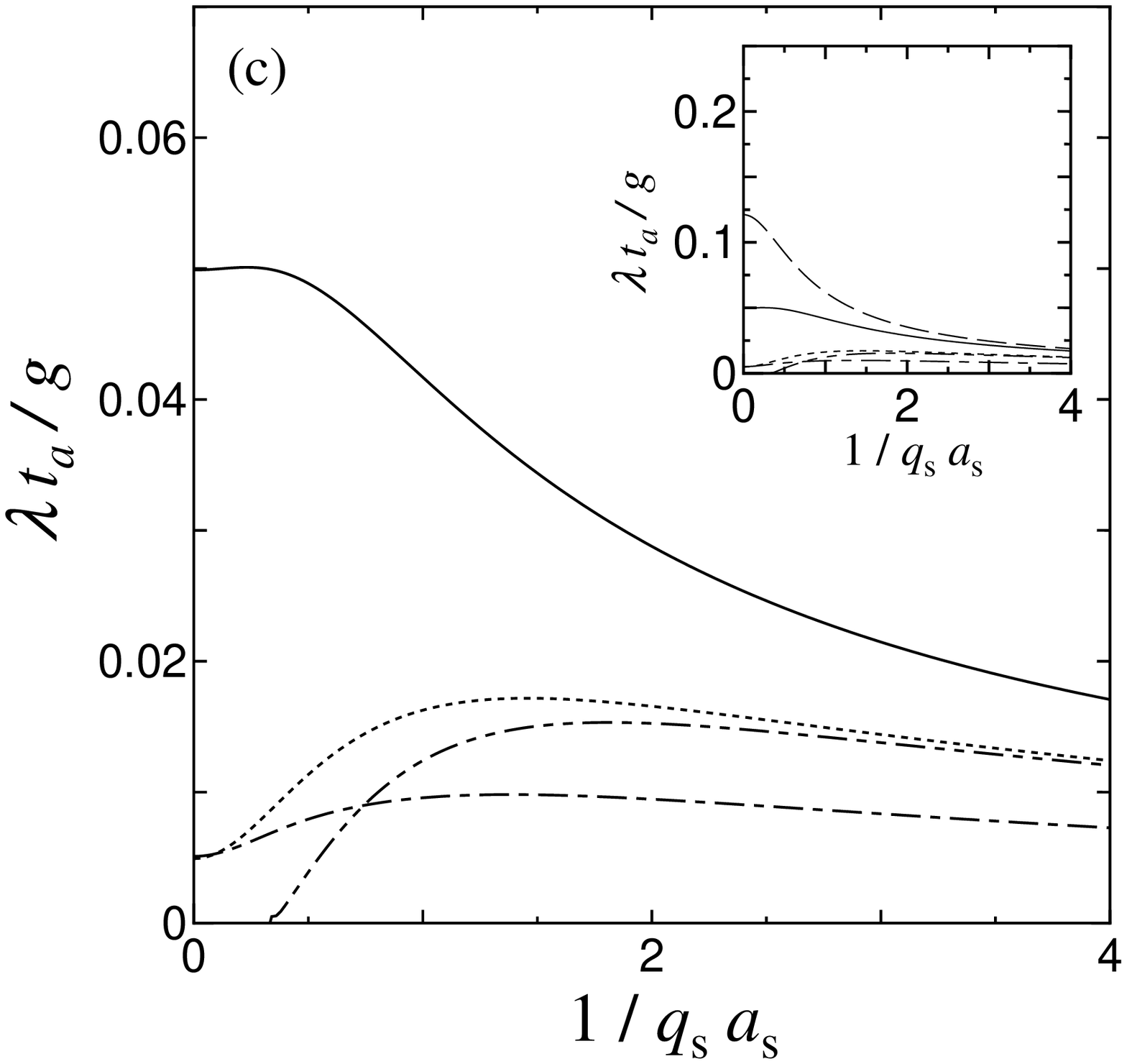}
\end{tabular}
\caption{\label{fig:01} Dimensionless coupling constants $\lambda$ 
                as functions of the screening length $\qs^{-1}$.
                The labels (a), (b) and (c) correspond 
                to the names of the models.
                The solid, dotted, dot-dashed and 2-dot-dashed curves 
                show the coupling constants $\lambda$ for 
                $p_x$, $p_y$, \textit{d} and $d_{xy}$-wave pairing, 
                respectively.
                In the inset, 
                the dashed curve shows the result for the \sw\ state.
                }
\end{center}
\end{fullfigure}

\begin{fullfigure}[t]
\begin{center}
\begin{tabular}{ccc}
\includegraphics[width=54mm,keepaspectratio]{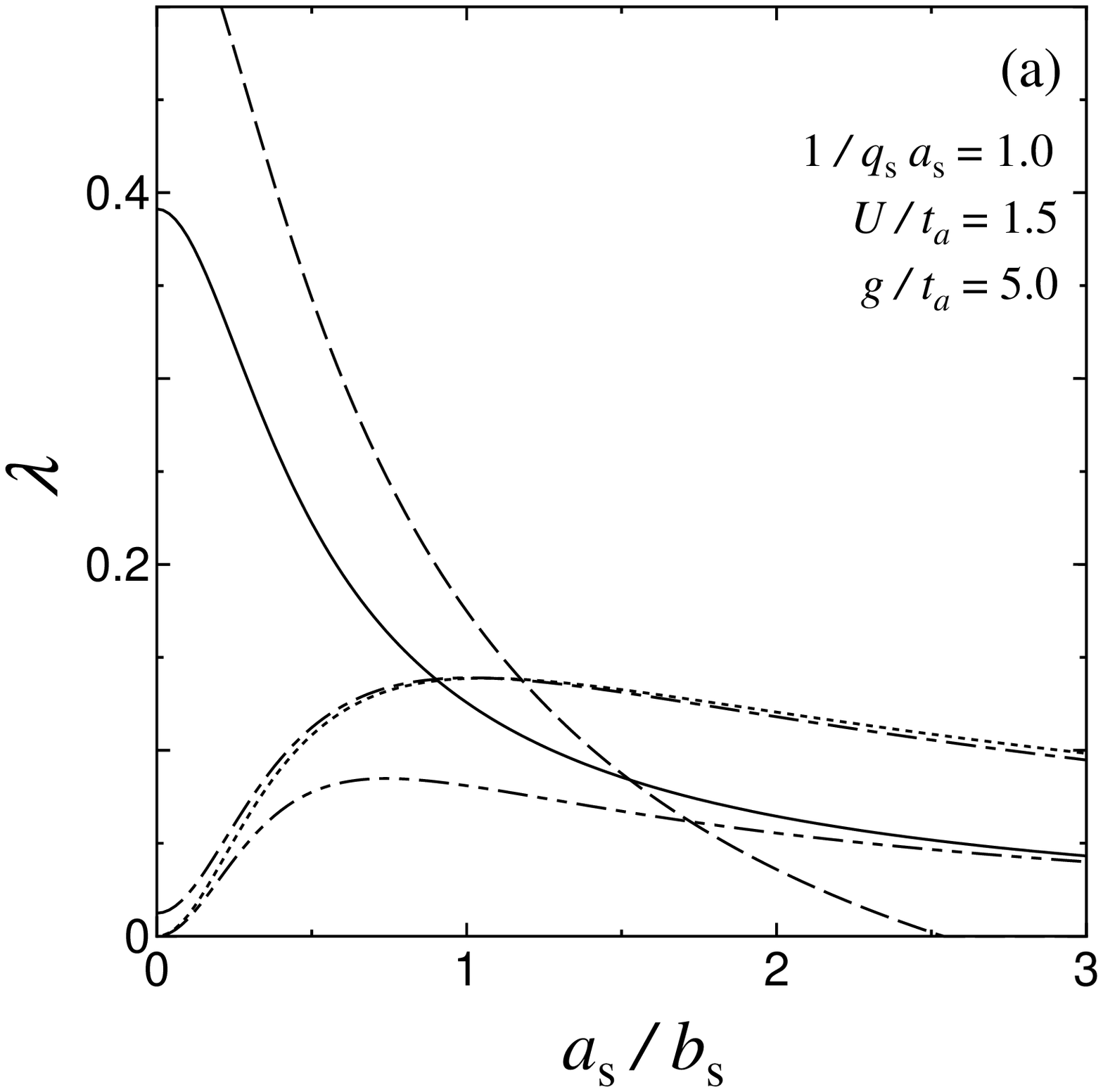} &
\includegraphics[width=54mm,keepaspectratio]{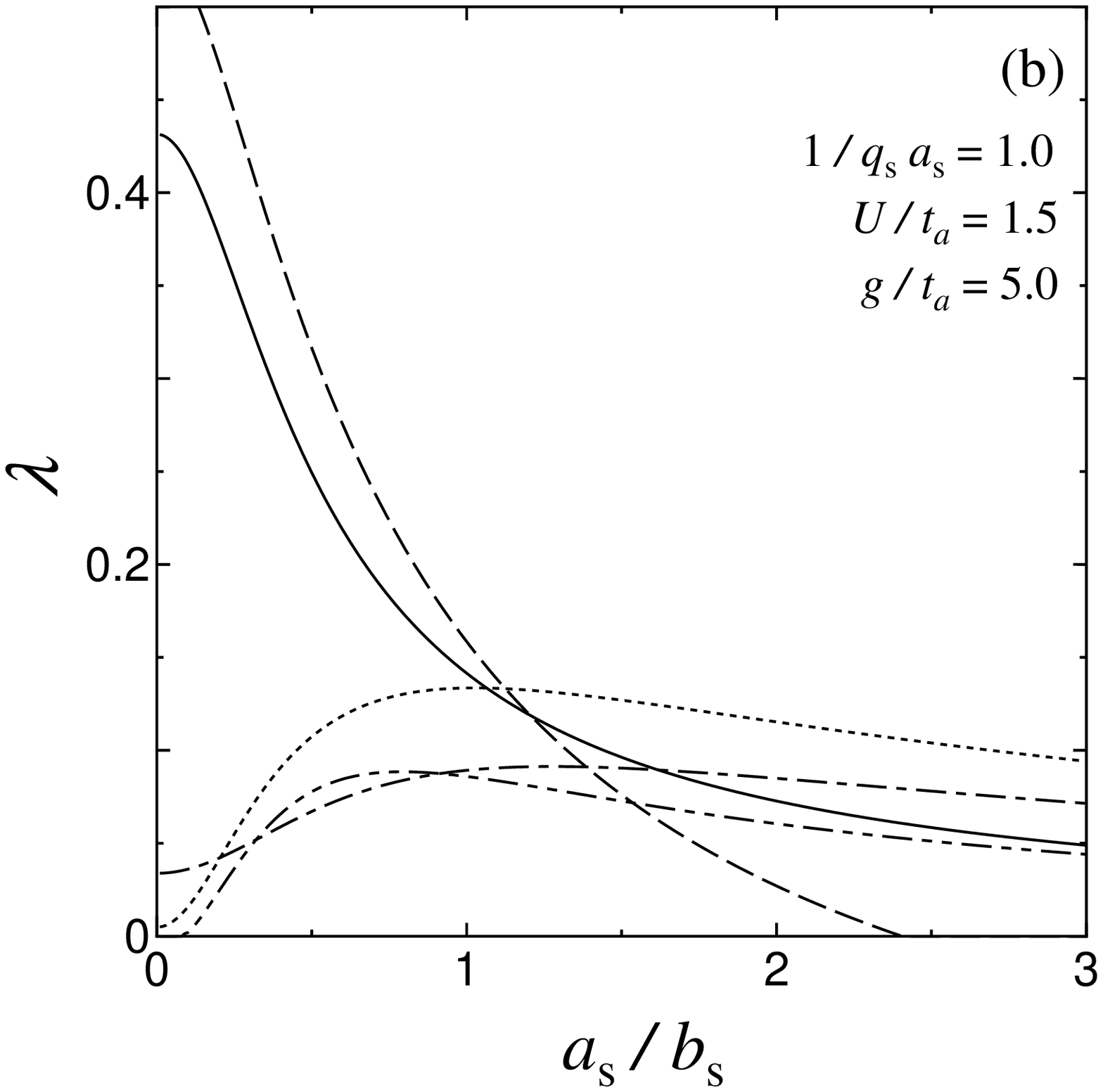} &
\includegraphics[width=54mm,keepaspectratio]{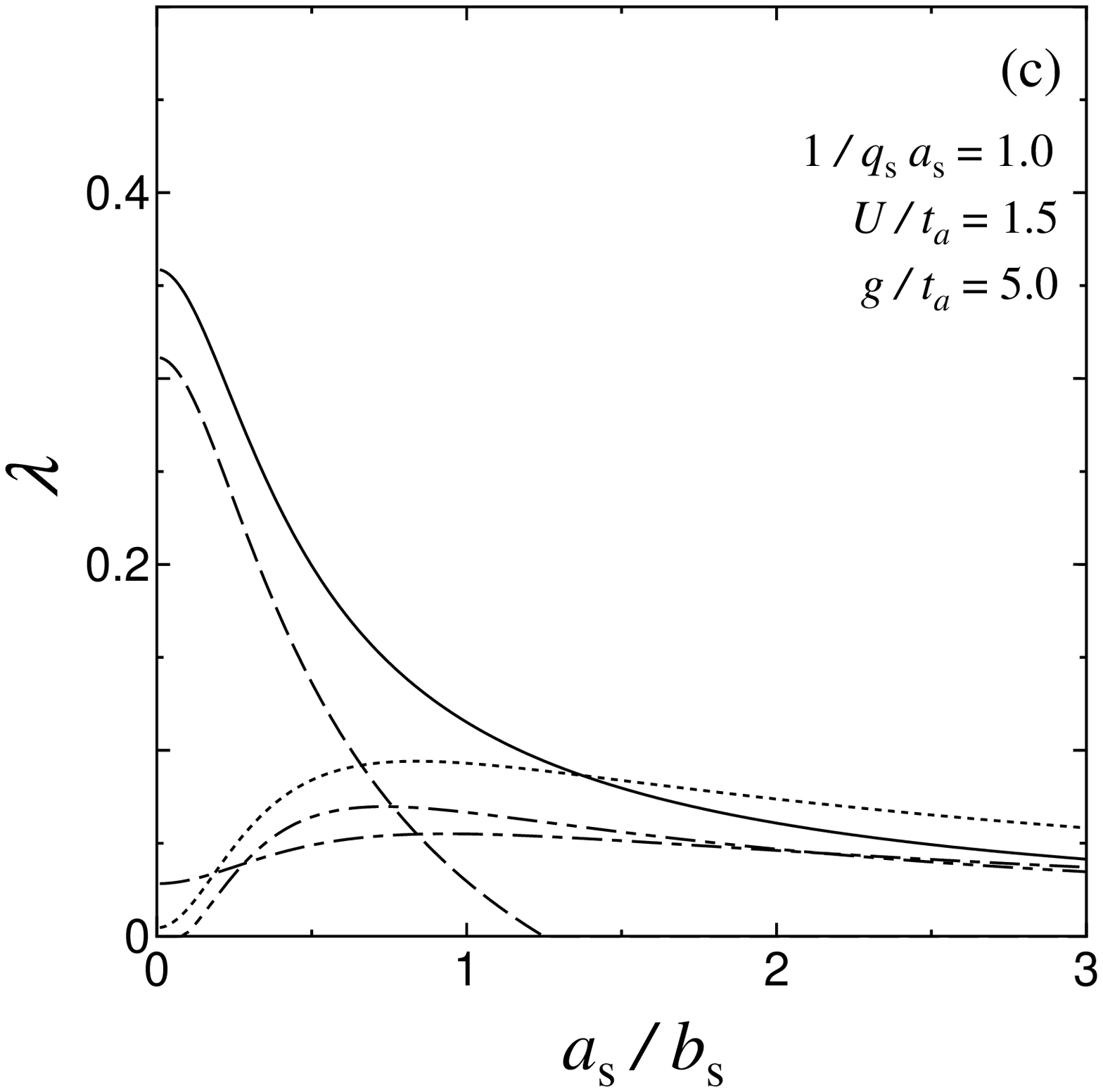}
\end{tabular}
\caption{\label{fig:02} Dimensionless coupling constants $\lambda$ 
                as functions of the ratio $\asbs$.
                The labels (a), (b) and (c) correspond 
                to the names of the models.
                The solid, dotted, dashed, dot-dashed and 2-dot-dashed curves 
                represent the results of $\lambda$ 
                for $p_x$, $p_y$, \textit{s}, \textit{d} 
                and $d_{xy}$-wave pairing, 
                respectively. 
                The Coulomb parameter $\mu_{\mathrm{C}}^\ast$ 
                is subtracted from $\lambda$ of \sw\ pairing.
                }
\end{center}
\end{fullfigure}

\begin{figure}[t]
\begin{center}
\includegraphics[width=65mm,keepaspectratio]{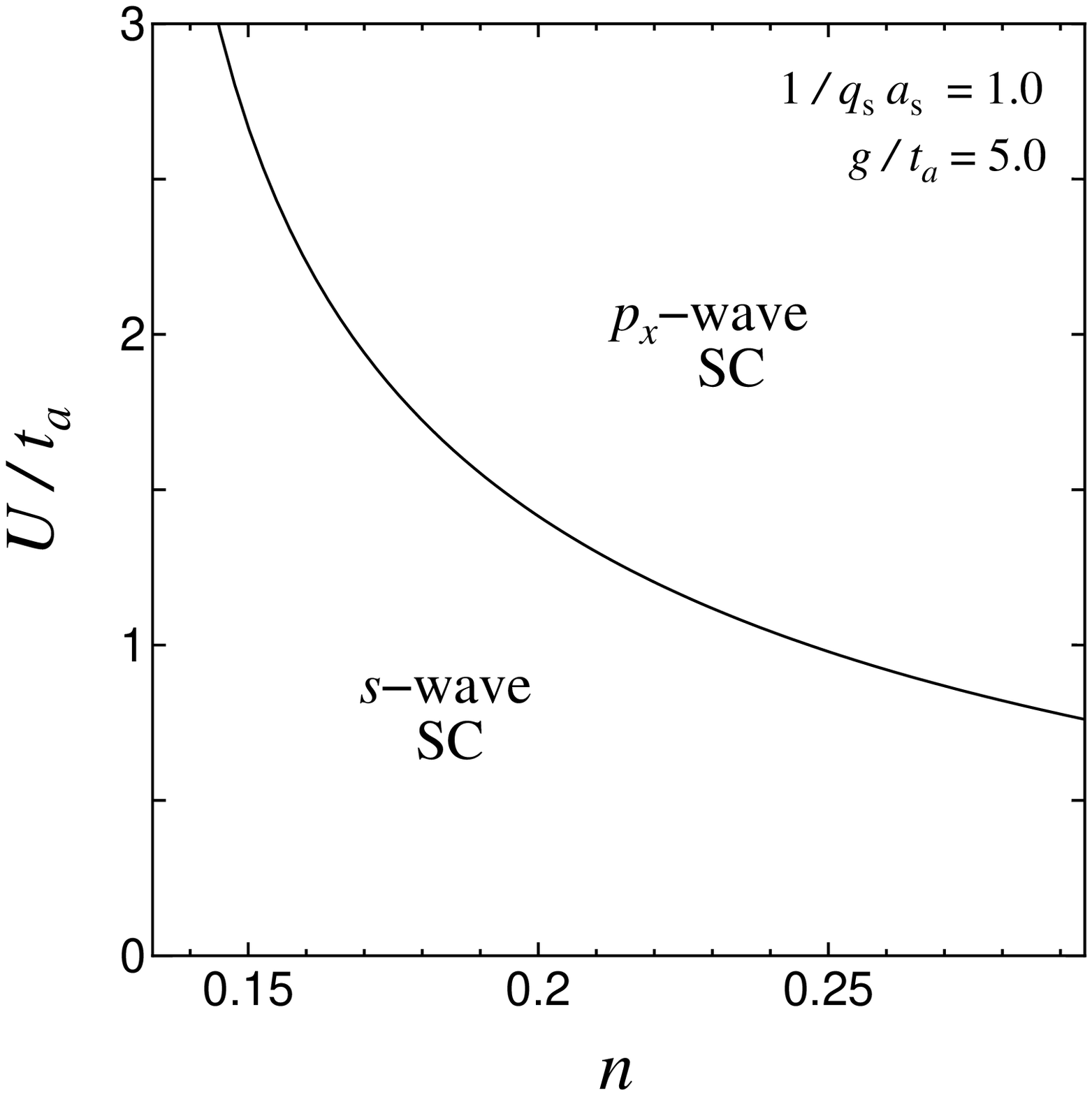}
\caption{\label{fig:05} Phase diagram in the $n$-$U$ plane at $T=0$. 
                }
\end{center}
\end{figure}

\begin{figure}[t]
\begin{center}
\includegraphics[width=65mm,keepaspectratio]{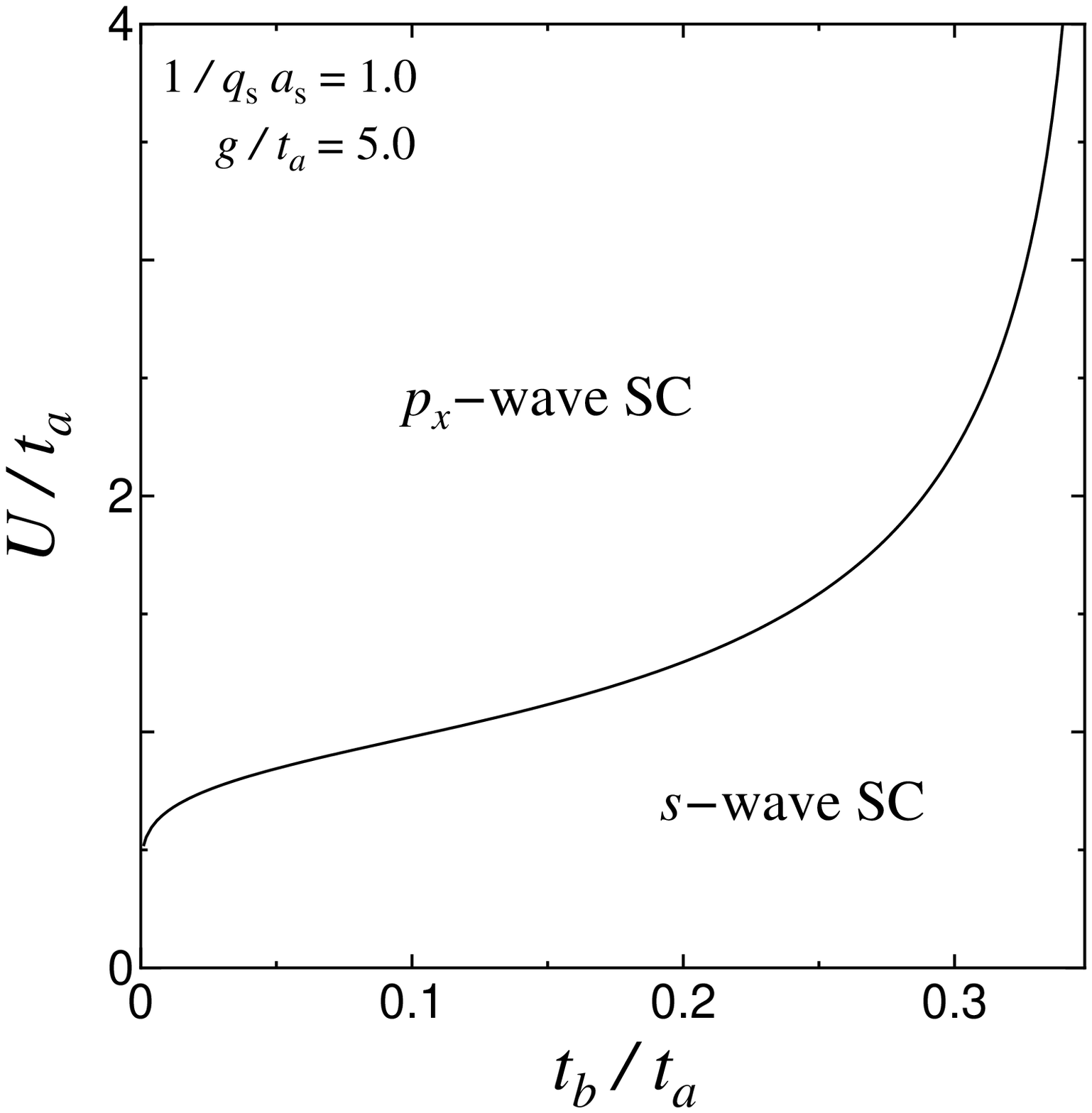}
\caption{\label{fig:06} Phase diagram in the $t_b$-$U$ plane at $T=0$. 
                }
\end{center}
\end{figure}

\begin{figure}[t]
\begin{center}
\includegraphics[width=65mm,keepaspectratio]{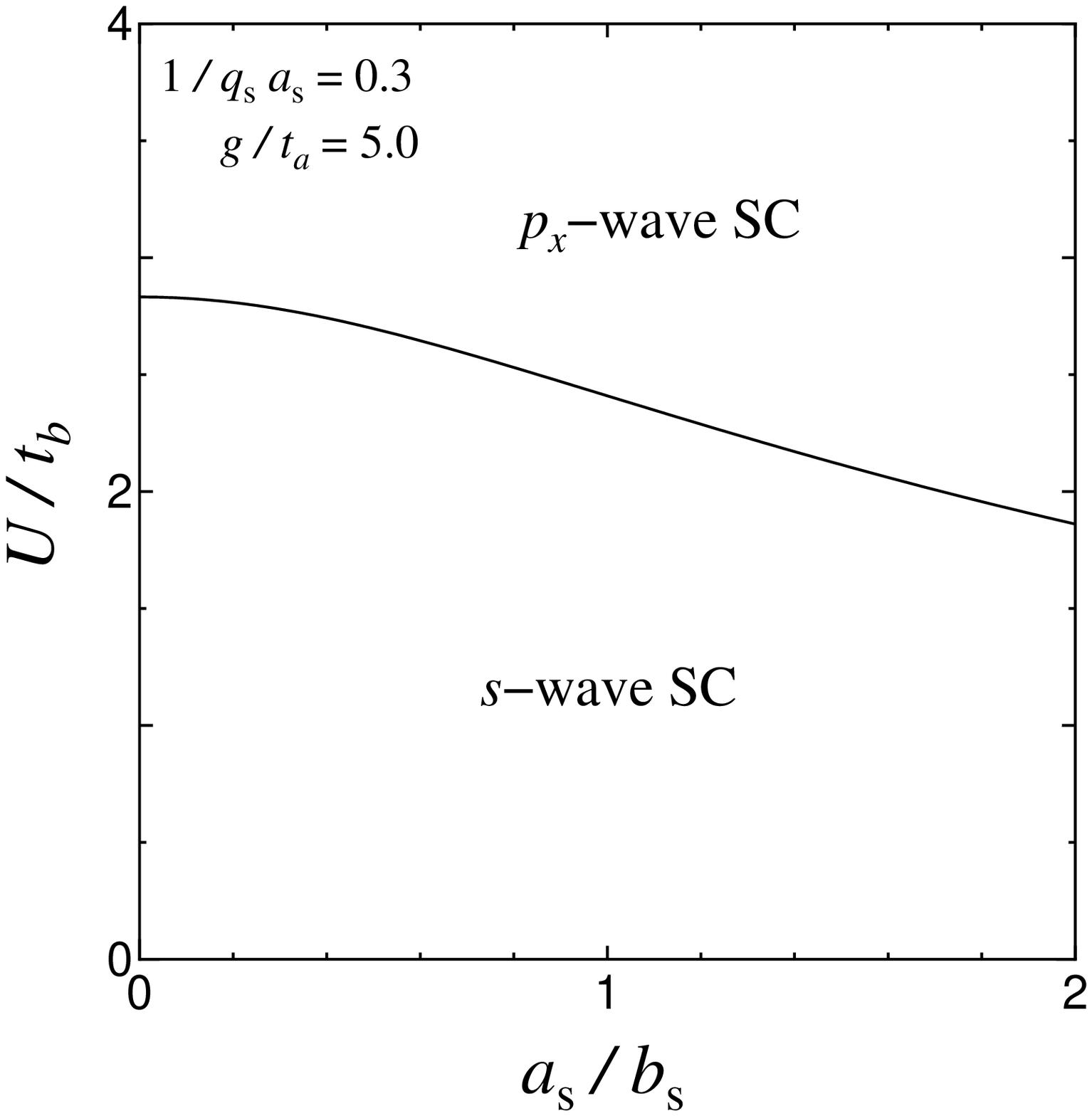} \\
\vspace{7mm}
\includegraphics[width=65mm,keepaspectratio]{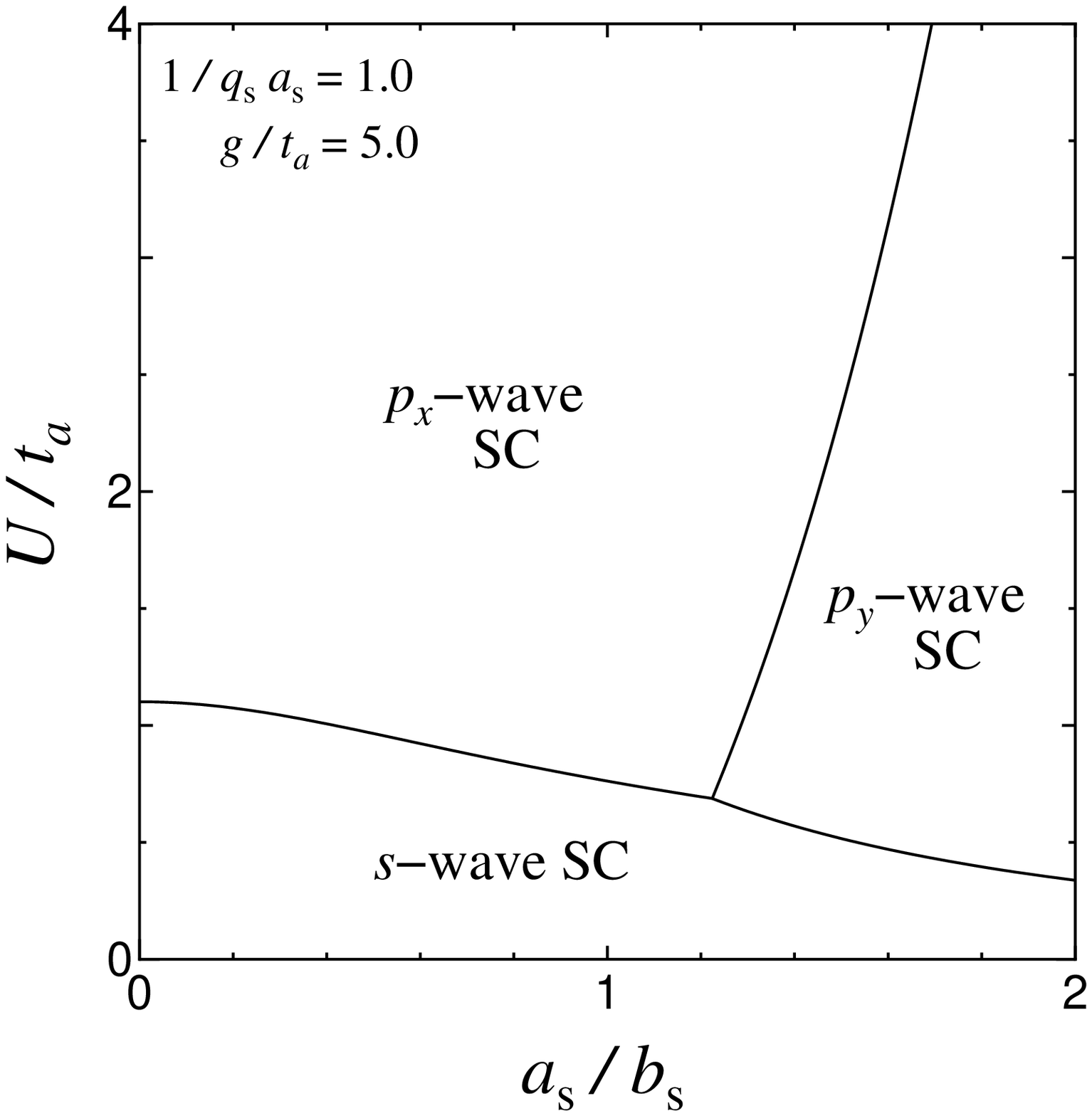}
\caption{\label{fig:07} Phase diagrams in the $\asbs$-$U$ plane at $T=0$. 
                }
\end{center}
\end{figure}

\begin{figure}[t]
\begin{center}
\includegraphics[width=65mm,keepaspectratio]{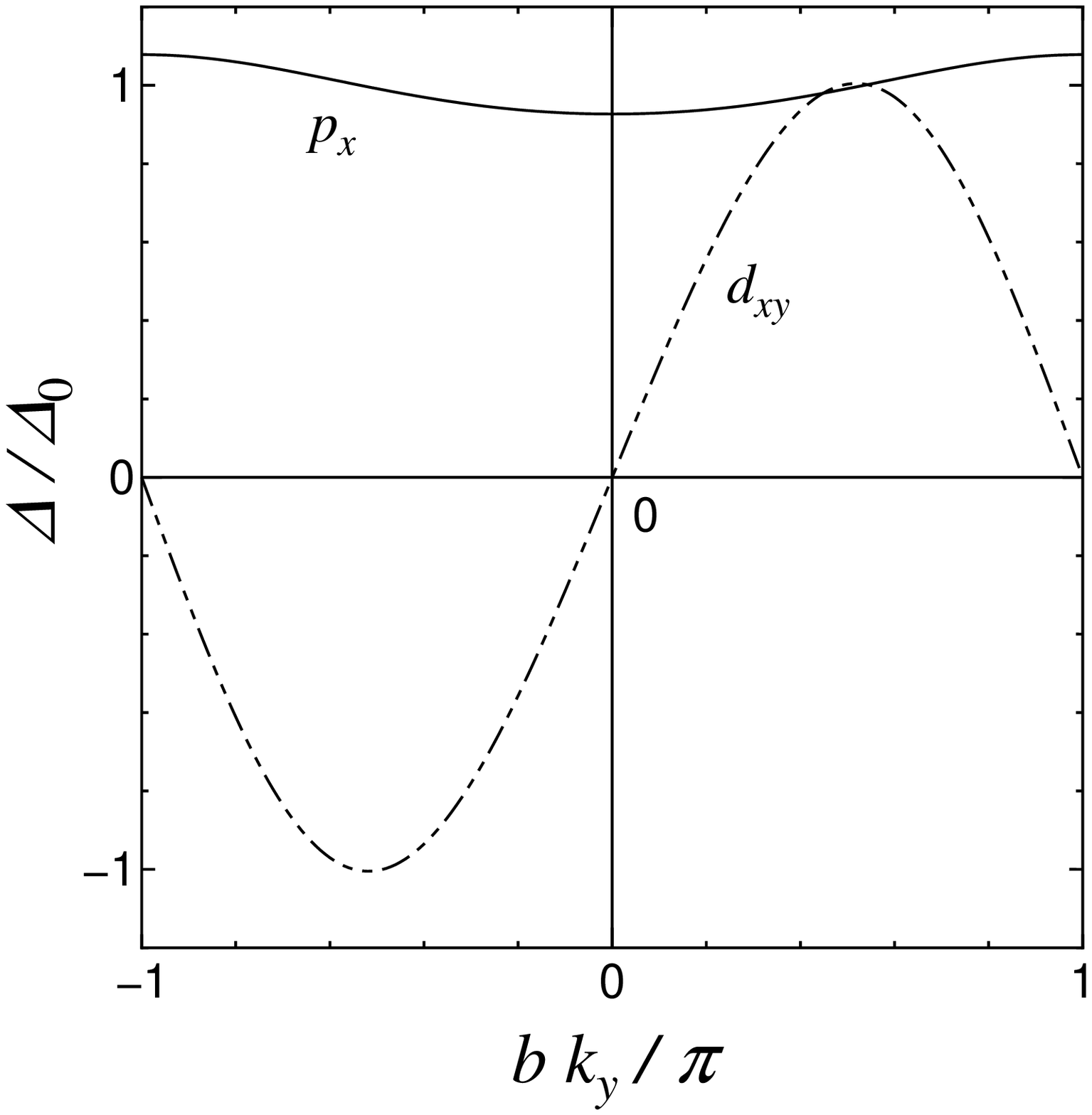}
\caption{\label{fig:09} Momentum dependences of the order parameters 
               $\tilde{\Delta}^{\mathrm{a}}(k_y) 
               = \sqrt{\rho(k_y)}\, \Delta(k_{\mathrm{F}x}(k_y), k_y)$ 
               of the $p_x$ and $d_{xy}$-wave states. 
               In these states, $\Delta(k_x, k_y) = - \Delta(-k_x, k_y)$.
                }
\end{center}
\end{figure}

In this section, 
we show the results of the numerical calculations.
In Figs.~\ref{fig:01}-\ref{fig:09}, 
we set $\omD = 200$~K~\cite{Yang1}, 
$t_b / t_a = 0.1$ (except for Fig.~\ref{fig:06}), 
$n = 1 / 4$ (except for Fig.~\ref{fig:05})
and $\asbs = 0.468$ 
(except for Figs.~\ref{fig:02} and \ref{fig:07}), 
considering the compounds \tmtx.
These compounds have a slight strain of the lattice, 
but we have confirmed 
by the numerical calculations
that it changes 
the result only very slightly. 
Therefore, 
we only show the results 
when the crystal axes are orthogonal.
We set $W_{\mathrm{C}} / \omD = 25.12$ 
in $\mu_{\mathrm{C}}^\ast$ from the experimental values~\cite{Yang1}, 
since $W_{\mathrm{C}} = \sqrt{(W - \mu')\mu'} \approx \sqrt{3}t_a$ 
for 1/4-filled band, 
where $W$ and $\mu'$ denote 
the band width 
and the chemical potential measured from the bottom of the band, 
respectively~\cite{Shimahara5,Ishiguro1}.

In Fig.~\ref{fig:01}, 
the results of 
the dimensionless coupling constants $\lambda$ are shown 
as functions of the screening length $\qs^{-1}$.
The labels (a)-(c) of figures correspond to the name of the models, 
which have been defined in the previous section.
The \sw\ coupling constant decreases rapidly 
with increase of the screening length $(\qs \as)^{-1}$ 
and is close to anisotropic coupling constants 
in $(\qs \as)^{-1} \gtrsim 1$.
In a whole region of $(\qs \as)^{-1}$, 
the \pxw\ state is more favorable than the \pyw\ state.
Especially for large $(\qs \as)^{-1}$, 
the \pxw\ state can easily overcome the \sw\ state 
with an assistance of $\mu_{\mathrm{C}}^\ast$.
We find the similar behavior 
in models~(a)-(c) for $(\qs \as)^{-1} \gtrsim 1$.

Figure \ref{fig:02} shows 
the dimensionless coupling constants $\lambda$ 
as functions of the ratio $\asbs$.
We assume 
$(\qs \as)^{-1} = 1$, 
$U / t_a = 1.5$ and $g / t_a = 5$ as an example. 
The \pyw\ state is favored for $\asbs \gtrsim 1$, 
while the \pxw\ state is favored for $\asbs \lesssim 1$, 
when \sw\ state is suppressed. 
In Figs.~\ref{fig:01} and \ref{fig:02}, 
it is also found by comparing the results of models~(a)-(c) 
that the \sw\ pairing interaction is weakened 
by the corrections due to the charge fluctuation 
and the short-range Coulomb interaction 
in addition to $\mu_{\mathrm{C}}^\ast$

\begin{figure}[t]
\begin{center}
\includegraphics[width=65mm,keepaspectratio]{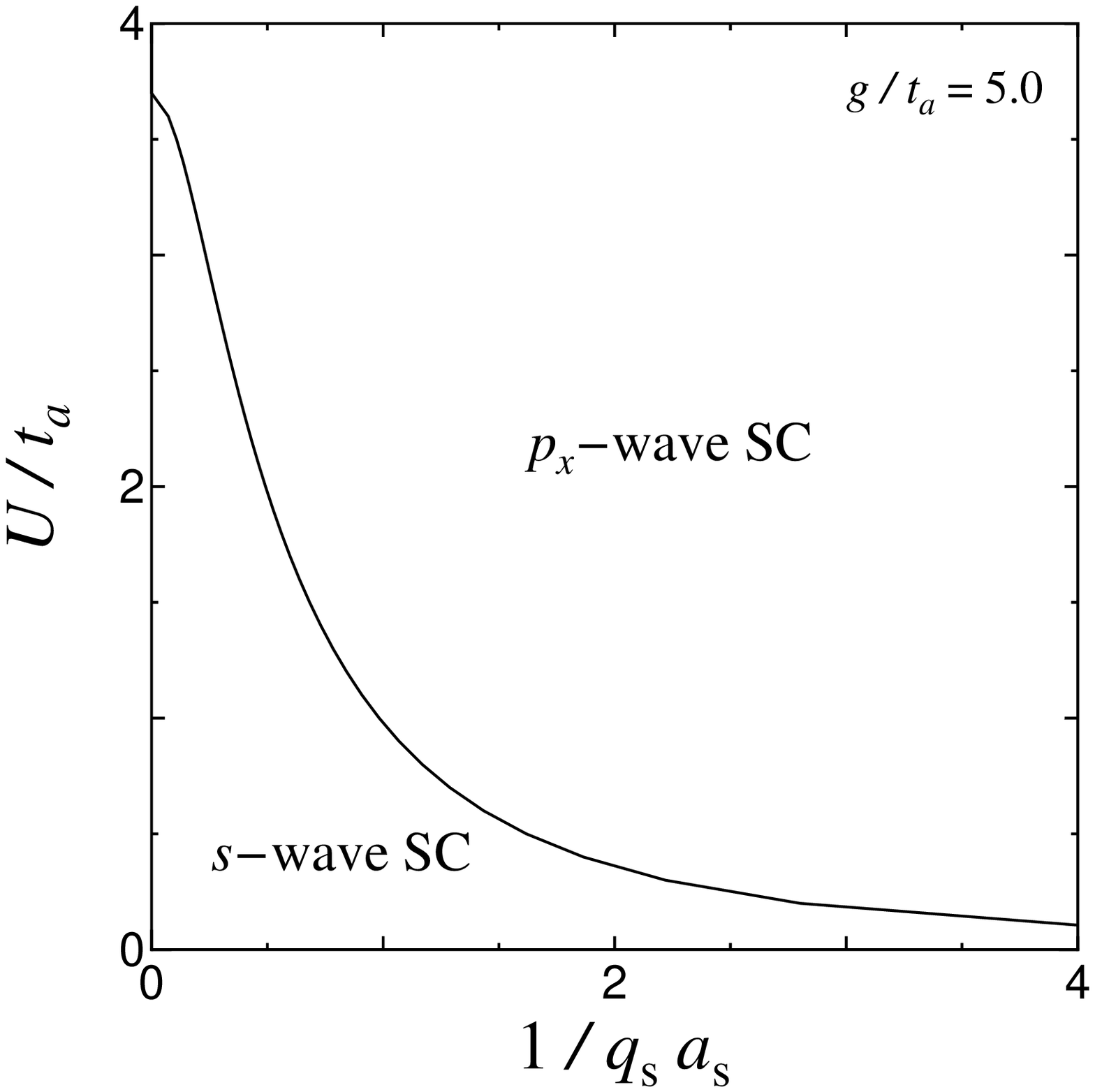}
\caption{\label{fig:03} Phase diagram 
                in the $\qs^{-1}$-$U$ plane at $T=0$. 
                Here, SC stands for superconductivity. 
                }
\end{center}
\end{figure}

\begin{figure}[t]
\begin{center}
\includegraphics[width=65mm,keepaspectratio]{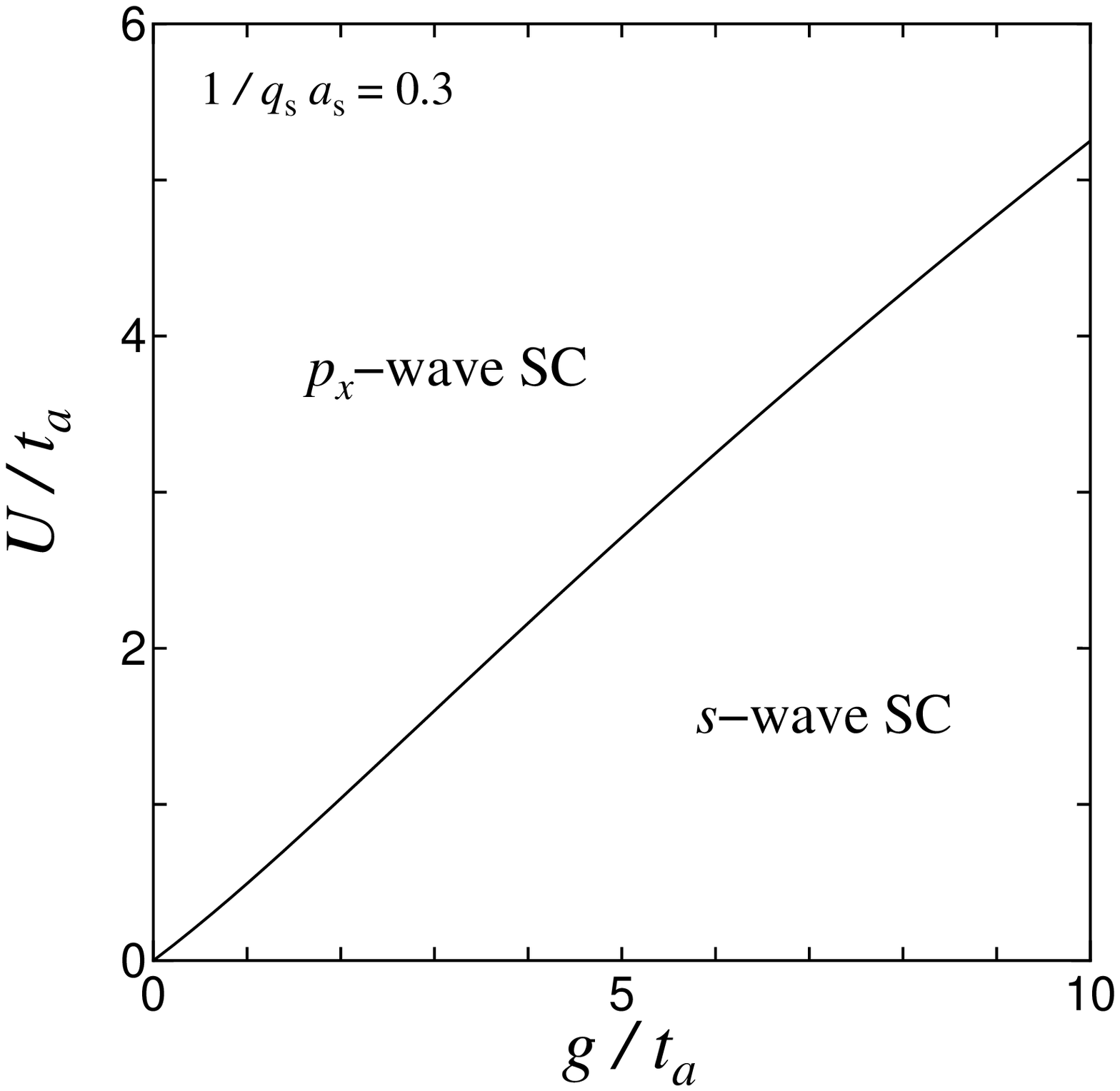} \\
\vspace{7mm}
\includegraphics[width=65mm,keepaspectratio]{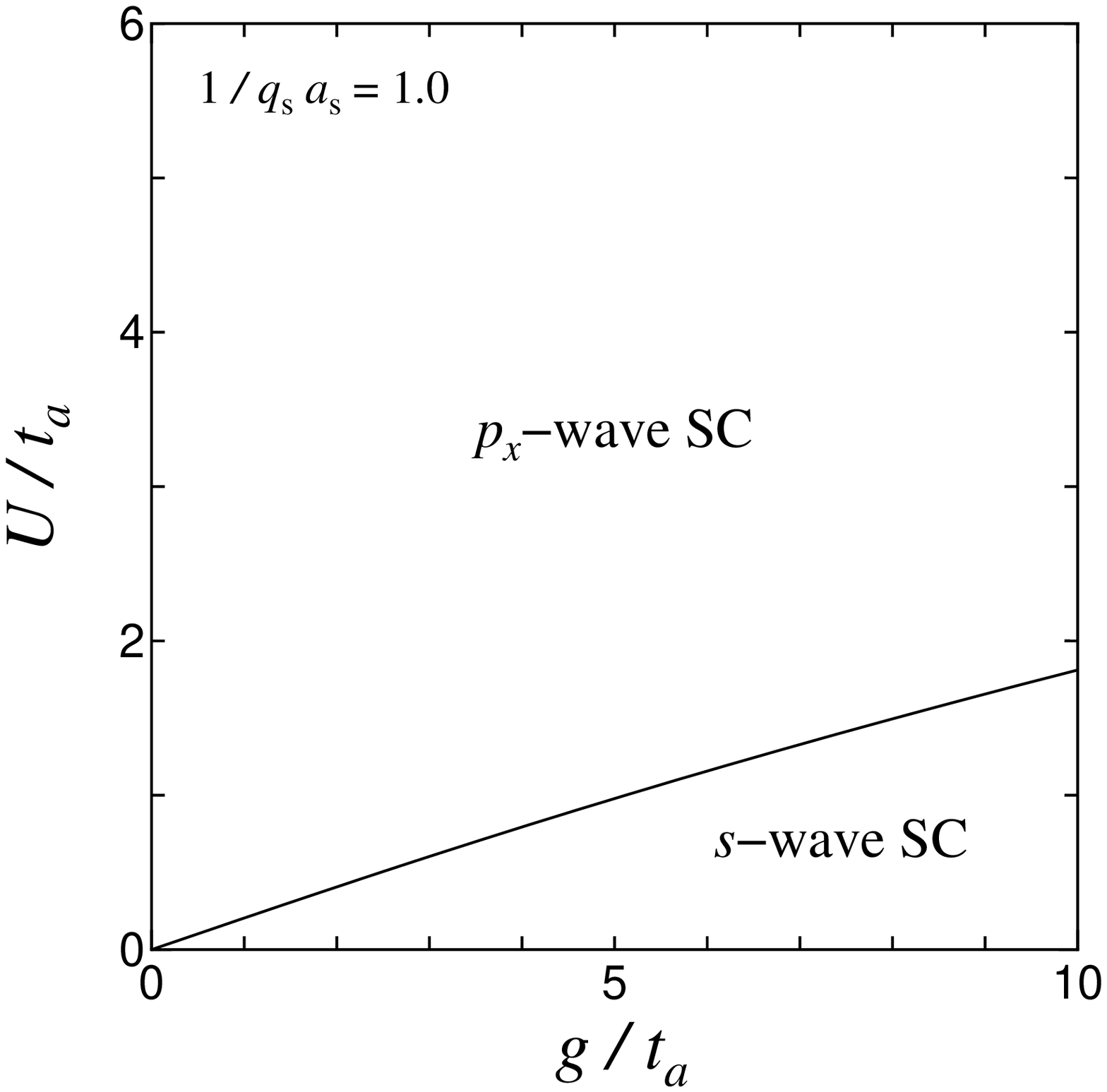}
\caption{\label{fig:04}Phase diagrams in the $g$-$U$ plane at $T=0$. 
                }
\end{center}
\end{figure}

\begin{figure}[t]
\begin{center}
\includegraphics[width=65mm,keepaspectratio]{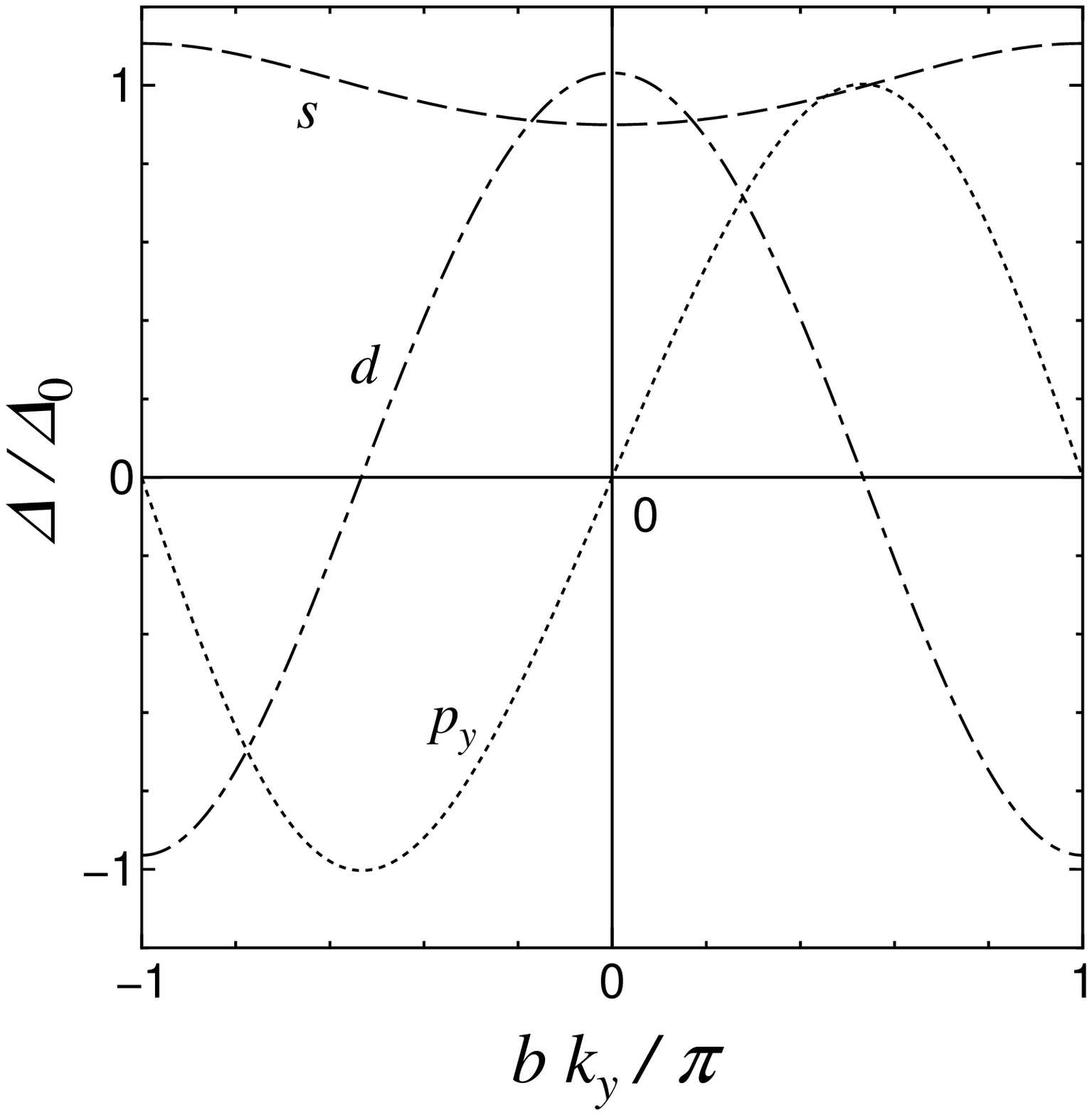}
\caption{\label{fig:08} Momentum dependences of the order parameters 
                $\tilde{\Delta}^{\mathrm{s}}(k_y) 
                = \sqrt{\rho(k_y)}\, \Delta(k_{\mathrm{F}x}(k_y), k_y)$ 
                of the \textit{s}, $p_y$ and \dw\ states. 
                In these states, $\Delta(k_x, k_y) = \Delta(-k_x, k_y)$. 
                }
\end{center}
\end{figure}

In Figs.~\ref{fig:03}-\ref{fig:09}, 
we concentrate ourselves to model~(c).
The results are qualitatively 
the same as those in models~(a) and (b).

Figure \ref{fig:03} shows
the phase diagram at $T = 0$ 
in the $(\qs \as)^{-1} $ - $ U / t_a$ plane. 
The parameter $g / t_a = 5$ is taken as an example.
It is found that the \pxw\ superconductivity occurs in the region 
where the screening effect is weak 
and the short-range interaction $U$ is strong.

Figure \ref{fig:04} shows
the phase diagrams at $T = 0$ 
in the $g / t_a $ - $ U / t_a$ plane. 
In each phase diagram, 
as is printed in it, 
the parameter $(\qs \as)^{-1} = 0.3$, 1 are taken as examples.
It is found that the \pxw\ superconductivity is favored 
for weak electron-phonon coupling 
and weak Coulomb screening.

Figure \ref{fig:05} shows the phase diagram at $T=0$ 
in the $n $ - $ U / t_a$ plane.
We show the results in the region of $n$ where the Fermi surface is open.
We set 
$(\qs \as)^{-1} = 1$ and $g / t_a = 5$ as an example. 
The \pxw\ superconductivity is favored for large $n$.

Figure \ref{fig:06} shows
the phase diagram at $T = 0$ 
in the $t_b / t_a $ - $ U / t_a$ plane. 
We set 
$(\qs \as)^{-1} = 1$ and $g / t_a = 5$ as an example. 
It is found that when $t_b / t_a$ increases, 
the \pxw\ superconductivity is suppressed.
Figure \ref{fig:07} shows
the phase diagrams at $T = 0$ 
in the $ \asbs $ - $ U / t_a$ plane. 
The parameters, $g / t_a = 5$, 
$(\qs \as)^{-1} = 0.3$ or 1 
are taken as examples.
It is found that the \pxw\ state is favored for small $\as$, 
while the \pyw\ state is favored for large $\as$.

These results are physically interpreted as follows.
The phonon-mediated pairing interactions 
are attractive in the whole momentum space of $\bq = \bk - \bk'$, 
and have a peak at $\bq = 0$.
Thus, 
the order parameter $\Delta(\bk)$ at $\bk$ is favorable 
if $\Delta(\bk')$'s of $\bk' \approx \bk$ 
have the same sign as $\Delta(\bk)$, 
while unfavorable 
if $\Delta(\bk')$'s at $\bk' \approx \bk$ 
have the opposite sign as $\Delta(\bk)$.
Therefore, 
the \pyw\ state is originally more unfavorable 
than the \pxw\ state, 
because of the node at $k_y = 0$, 
where $\Delta(\bk)$ changes its sign.
However, 
when the areas of the Fermi surfaces of 
$k_x > 0$ and $k_x < 0$ approach 
by increasing $\as$, 
the \pxw\ state becomes unfavorable, 
because $\Delta(k_x > 0, k_y)$ 
and $\Delta(k_x < 0, k_y)$ have the opposite signs.
The dependences on $t_b$ of the coupling constants are also explained 
in the same manner as 
the dependences on the ratio $\asbs$.
Near the ends of the first Brillouin zone, 
the distance between the sheets of the Fermi surface 
becomes narrower as $t_b$ increase.
(For $t_b /t_a \gtrsim 0.35$, 
the Fermi surface is closed.)
For the \pxw\ state, 
the interaction becomes repulsive 
for pair hopping $\Delta(\bk) \rightarrow \Delta(\bk')$ 
near the zone ends.
As a result, 
as $t_b$ increases, 
\sw\ superconductivity becomes more favorable than \pxw\ one.

Figures \ref{fig:08} and \ref{fig:09} 
show the calculated momentum dependences of the order parameters.
We assume 
$(\qs \as)^{-1} = 1$ and 
$U / t_a = 1.5$ as an example in both figures. 
It is confirmed that 
the $p_y$ and $d_{xy}$-wave states have a line node 
on the Fermi surface at $k_y =0$, 
the \dw\ state has line nodes at $k_y \approx \pm \pi / 2 b$, 
and the \textit{s} and \pxw\ states are full gap states.

\section{\label{sec:sum}Summary and Discussion}

Now, 
we summarize the results.
We have examined pairing interactions 
mediated by phonons 
in quasi-one-dimensional (Q1D) systems 
with open Fermi surfaces.
It was found that spin-triplet superconductivity 
occurs for weak Coulomb screening 
and strong on-site Coulomb interaction $U$, 
even in the absence of any additional nonphonon 
pairing interactions.
We have examined two possible spin-triplet states.
One has a nodeless gap function $\Delta(\bk)$ 
for the open Fermi surface, 
while another has line nodes 
at $k_y = 0$ on the Fermi surface.
We call the former \pxw\ state, 
while the latter \pyw\ state.
We have found that 
when the Fermi surface is open, 
as the ratio $\asbs$ increase, 
\pyw\ superconductivity becomes more favorable.
We have explained the physical reason for it. 
It was argued that 
the \pxw\ state occurs when 
the conductive chains are separated with a sufficiently large spacing.

In \tmtx, 
the ratio $\asbs = 0.468$ 
and the hole density $n = 1 / 4$ 
give sufficiently large separation of the
$k_x > 0$ and $k_x < 0$ sheets of the open Fermi surface 
when $t_b = 0.1 t_a$.
Hence, 
the \pxw\ state is more favorable 
than the \pyw\ state 
in those compounds.
It was also found 
that the \pxw\ state is more favorable
than the \sw\ state for $n = 1 / 4$ and $U \approx 1.5 t_a$, 
which are appropriate to \tmtx, 
if we set $(\qs \as)^{-1} \gtrsim 1$, 
which is physically plausible.

In our model, 
$t_b$ may be regarded as an effective parameter 
that reflects an effect of pressure.
If $\asbs$ and $t_b$ increase with increase of the pressure, 
the coupling constant for \pxw\ pairing decreases 
as found in Figs.~\ref{fig:02} and \ref{fig:06}.
This pressure dependence may contribute 
to the observed pressure dependence of $\Tc$ to some extent.

In order to discuss 
the reality of the phonon-mediated spin-triplet superconductivity, 
we crudely estimate the parameters 
for \tmtx\ from the observed transition temperature 
$\Tc \simeq 1$~K.
Here, 
we assume \pxw\ pairing.
If we assume 
$\omD \simeq 200$~K~\cite{Yang1} and $\Tc \simeq 1$~K, 
we have $\lambda \simeq 0.184$.
Then we find $g \simeq 4.61 t_a$ 
for $(\qs \as)^{-1} \simeq 1$ 
from Fig.~\ref{fig:01}.
For such a choice of parameter values, 
in order to suppress 
\sw\ pairing, 
the on-site Coulomb repulsion 
must be $U \gtrsim 2 t_a \simeq 0.5 W$, 
which seems realistic as the order of the magnitude.
This value is rather larger 
than the estimation $U \approx 1.5 t_a$ 
from the SDW transition. 
However, 
it can be explained by taking into account that 
the repulsive on-site interaction is enhanced 
by the spin fluctuation near the SDW phase from the bare value $U$ 
in practice.

In conclusion, 
the nodeless spin-triplet superconductivity 
is favorable in the compounds \tmtx, 
if \sw\ superconductivity is suppressed.
The present result consistent 
with the experimental data of the Knight shift and thermal conductivity.
Moreover, 
it was also found that 
the following three conditions are 
indispensable for the appearance 
of the nodeless spin-triplet superconductivity 
mediated by phonons: 
(1) The screening effect 
    is sufficiently weak; 
(2) The short-range Coulomb interaction is strong; 
(3) The distance between sheets of the open Fermi surface 
    is sufficiently large. 
Here, 
conditions (1) and (2) 
are satisfied in \tmtx, 
in which it is known that the electron states 
are well-described by the tight-binding model, 
and the layer spacing 
is large.
The condition (3) is satisfied 
by the quasi-one-dimensionality ($\asbs = 0.468$)
and the hole density $n = 1/4$.

In \tmtx, the pressure-temperature phase diagrams 
of superconductivity and SDW were obtained~\cite{Jerome0}.
The pressure dependence may be reproduced 
by taking into account the pairing interactions 
mediated by antiferromagnetic fluctuations, 
since those interactions have 
attractive triplet components~\cite{Shimahara1,Shimahara8}.
We have shown that the \pxw\ interaction 
mediated by phonons also decreases 
to some extent as the pressure increases.
The explanation of the experimental phase diagrams 
in the present theory 
including the effect of the spin fluctuations
remains for a future study.

\section*{Acknowledgment}

This work was partly supported by a Grant-in-Aid 
for COE Research (No.13CE2002) 
of the Ministry of Education, 
Culture, Sports, Science and Technology of Japan.


\begin{thebibliography}{99}
\bibitem{footnote2}
In p.~272 of ref.~\citen{Ishiguro1}, 
it is claimed that since $T_{\mathrm{SDW}} = 25$~K 
is assumed at $t_b = 0$ in ref.~\citen{Shimahara1}, 
the obtained $T_c \lesssim 1.0$~K is slightly too low. 
However, in actuality, 
$T_{\mathrm{SDW}} = 20$~K is assumed at $t_b = 0$ 
in ref.~\citen{Shimahara1}, 
and it gives $T_{\mathrm{SDW}} \simeq 12$~K 
at $t_b / t_a \approx 0.135 \sim 0.1 $, 
which agrees with the experimental facts, 
as the order of magnitude.
\bibitem{footnote1}
In actuality, 
the unit cell volume of \tmtp\ is equal to $714.3$~\AA$^3$. 
( See ref.~\citen{Ishiguro1}.) 
However, 
since the anions are far from the conduction layers of TMTSF molecules, 
we regard that the unit cell consists of a single TMTSF molecule.
Therefore, 
in our model, we set $\Vc = 714.3/2 = 357.2$~\AA$^3$.
\end{thebibliography}
\end{document}